\documentclass[review]{elsarticle}

\usepackage{lineno,hyperref}
\usepackage{amssymb}
\usepackage{graphicx}
\usepackage[ruled,vlined,linesnumbered]{algorithm2e}
\usepackage{float}
\usepackage{color}
\usepackage{soul}
\usepackage{amsmath}
\usepackage{caption}
\usepackage{makecell}
\usepackage{enumitem} 
\usepackage{makecell}
\usepackage{subcaption}
\modulolinenumbers[5]

\journal{Journal: Internet of Things: Engineering Cyber-Physical Human Systems (Elsevier),}

\begin{document} 

\begin{frontmatter}

\title{AppsPred: Predicting Context-Aware Smartphone Apps using Random Forest Learning}

\author[a,b]{Iqbal H. Sarker$^*$}
\cortext[mycorrespondingauthor]{Corresponding author}
\ead{msarker@swin.edu.au}
\author[c]{Khaled Salah}

\address[a]{Department of Computer Science and Software Engineering, \\Swinburne University of Technology, \\ Melbourne, VIC-3122, Australia.}
\address[b]{Department of Computer Science and Engineering, \\Chittagong University of Engineering and Technology, Bangladesh.}
\address[c]{Department of Electrical and Computer Engineering, \\ Khalifa University, UAE.}

\begin{abstract}
Due to the popularity of context-awareness in the Internet of Things (IoT) and the recent advanced features in the most popular IoT device, i.e., smartphone, modeling and predicting personalized usage behavior based on relevant contexts can be highly useful in assisting them to carry out daily routines and activities. Usage patterns of different categories smartphone apps such as social networking, communication, entertainment, or daily life services related apps usually vary greatly between individuals. People use these apps differently in different contexts, such as temporal context, spatial context, individual mood and preference, work status, Internet connectivity like Wifi status, or device related status like phone profile, battery level etc. Thus, we consider individuals' apps usage as a \textit{multi-class context-aware} problem for personalized modeling and prediction. Random Forest learning is one of the most popular machine learning techniques to build a multi-class prediction model. Therefore, in this paper, we present an effective context-aware smartphone apps prediction model, and name it \textit{``AppsPred''} using random forest \textit{machine learning} technique that takes into account optimal number of trees based on such multi-dimensional contexts to build the resultant forest. The effectiveness of this model is examined by conducting experiments on smartphone apps usage datasets collected from individual users. The experimental results show that our AppsPred significantly outperforms other popular machine learning classification approaches like ZeroR, Naive Bayes, Decision Tree, Support Vector Machines, Logistic Regression while predicting smartphone apps in various context-aware test cases.
\end{abstract}

\begin{keyword}
\texttt{Smartphones; machine learning; mobile data mining; apps usage modeling; predictive analytics; context-aware computing; IoT analytics; personalization; intelligent services;}
\end{keyword}

\end{frontmatter}

%\linenumbers

\section{Introduction}
Nowadays, smart mobile phones are considered as one of the most popular IoT devices and have become an essential part of our everyday life. In the real world, users' interest on ``Mobile Phones''
is more and more than other platforms like ``Desktop
Computer'' or ``Tablet Computer'' over time \cite{sarker2018MobileDataScience}.  According to ITU (International Telecommunication Union), cellular network coverage has reached 96.8\% of the world population, and this number even reaches 100\% of the population in the developed countries, like USA, Australia, Canada, UK etc. in the world \cite{number-of-mobile-phone-users}. People use smartphones for not only the voice communication between individuals but also a variety of applications, apps in short, for different purposes like social networking, instant messaging, location tracking and transportation management, online shopping, medical appointment or eHealth services, sports and entertainment, IoT services, or real life emergency services etc. Usage patterns of such category of apps usually vary greatly between individuals in the real world. Individual users may behave differently in different contexts, such as temporal context including their work status like workday or holiday, spatial context that represents user's particular location, e.g., office, their emotional state or mood, e.g., happy, Internet connectivity like Wifi status, or device related status like phone profile or battery level etc. in which that usage occurs. Thus, its important to study on such contextual data in order to build an effective context-aware apps prediction model.

Developing personalized context-aware methods to model different categories of smartphone apps, particularly, Gmail, Microsoft Outlook, Facebook, LinkedIn, Twitter, Youtube, Whatsapp, Skype, eHealth, Uber, Browser, Google Maps etc. utilizing contextual data is the key. Thus, we consider this issue as a multi-class context-aware prediction problem, where each individual app represents as a particular usage class. An effective machine learning based context-aware model by analyzing individuals' usage patterns in multi-dimensional contexts mentioned above utilizing smartphone data can eventually predict future usage according to their current contexts \cite{sarker2018research}. Such context-aware model can be used for building various data-driven intelligent systems, such as intelligent mobile recommendation system, context-aware smart apps management system, context-aware smart app searching, intelligent app notification management system etc. that intelligently assist the end mobile phone users in their daily activities \cite{sarker2018BehavMiner}. Therefore, in order to achieve our goal, in this paper, we mainly focus on modeling and predicting personalized \textit{smartphone apps usage} based on relevant multi-dimensional contexts related to the corresponding users' preferences and their own devices characteristics.

Let's consider a real-world motivational example. Say, Alice, a smartphone user, is a post graduate research student. She has installed a large number of mobile applications on her smartphone. Homescreens of smartphones provide easy finding of the apps without additional effort in searching, which is useful to the end mobile phone users in their various day-to-day situations. However, the homescreen of her smartphone is unaware about the current contexts of her. As a result, the phone becomes unable to manage the useful apps intelligently according to her needs, as her current contexts, e.g., location, are not static, may change over time. An effective context-aware app usage model may predict her future usage based on her current contexts, allowing the particular app she currently needs to be easily accessible from the mobile homescreen. Such personalized model could be used to build a smart mobile app management system that can predict her future usage according to her current contextual information and intelligently assists her to use different categories of smartphone apps according to her needs.

In the area of contextual smartphone data analytics, both association learning \cite{agrawal1994fast}, and classification learning \cite{quinlan1993} are the most common and popular techniques to build a data-driven prediction model. However, association learning technique, e.g., Apriori \cite{agrawal1994fast} produces a large number of redundant rules that makes the context-aware prediction model more complex and ineffective \cite{fournier2012mining} \cite{sarker2018mining}. Thus, in this paper, we focus on classification techniques that can play an important role to build an effective context-aware prediction model for individual mobile phone users utilizing their smartphone apps usage data. In the area of machine learning and predictive analytics, ZeroR, Naive Bayes, Decision Tree, Support Vector Machines, Logistic Regression, and Random Forest are the most popular classification algorithms that can be used to build data-driven context-aware models \cite{sarker2019classifications} \cite{han2011data}. Among these techniques, tree based context-aware model is more effective to intelligently predict mobile user activity in different contexts \cite{sarker2019classifications}. In particular, a number of researchers \cite{hong2009context} \cite{lee2007deploying} \cite{zulkernain2010mobile} \cite{sarker2019machine} have used decision tree classifier to model mobile phone users' behavior. Since we take into account our apps prediction model as a multi-class problem that includes a variety of usage classes in a number of multi-dimensional contexts, a single decision tree may cause over-fitting problem, while selecting the root node based on contexts. As a result, it may decrease the prediction accuracy of the resultant context-aware model. Thus, the research question is - \textit{How to build an effective context-aware smartphone apps usage prediction model for personalized services?}

In this paper, we present a \textit{random forest} machine learning based context-aware smartphone apps prediction model, \textit{``AppsPred''} that takes into account a number of trees rather than a single decision tree. In our model, we first extract the contextual features from the training dataset and prepare the contexts to fit for the machine learning techniques. Once the contexts have been processed, we then construct a random forest on the processed training dataset to achieve our goal. The reason for constructing random forest is that it averages the output of several separate learners like single decision tree by reducing the variance in individual's usage. However, different number of trees may give different prediction results in a random forest based model. Thus, in order to build an effective context-aware model, we take into account an \textit{optimal number of trees} that gives higher accuracy while predicting different categories of smartphone apps in different context-aware test cases. The effectiveness of this model is examined by considering the real mobile phone datasets consisting of individuals' various app usage and corresponding contextual information.

The contributions of this work can be summarized as follows. 

\begin{itemize}
	\item We first highlight the significance of personalized smartphone apps usage prediction modeling based on machine learning techniques. In our model, we take into account different categories of apps usages like social networking, communication, entertainment and so on in different multi-dimensional contexts related to the corresponding users' day-to-day situations and preferences, and their own devices' characteristics.
	
	\item We have collected contextual apps usage datasets form individual smartphone users and present a data-driven context-aware smartphone apps prediction model, ``AppsPred'' using random forest learning that takes into account an optimal number of trees based on relevant multi-dimensional contexts to make the model effective.
	
	\item Finally, we conduct experiments on the real-world collected datasets and evaluate the effectiveness of our AppsPred model for various context-aware test cases. The experimental results show that our AppsPred significantly outperforms other popular machine learning classification approaches.
\end{itemize}  

The rest of the paper is organized as follows. Section \ref{background} provides background and related work. In section \ref{Problem-Statement}, we define and formulate the problem addressing in this paper. In Section \ref{Methodology}, we present our context-aware smartphone apps usage prediction model using random forest machine learning. We report the experimental results in Section \ref{Evaluation}. We also summarize a number of key points in Section \ref{Discussion} and finally Section \ref{Conclusion} concludes this paper and highlights the future work.

\section{Background and Related Work}
\label{background}

In the area of contextual smartphone data analytics, both association learning and classification learning are the most common and popular machine learning techniques to build a prediction model. Association learning is the discovery of rules or patterns among a set of available items in a given dataset. Association learning technique, e.g., Apriori, is well defined in terms of the reliability and flexibility as it has the own parameters; the support, and the confidence \cite{agrawal1994fast}. It discovers association rules that satisfy the predefined minimum support and confidence constraints from a given dataset. Support of a rule represents the percentage of records in the dataset which carry all the items or contexts in a rule, and the confidence represents the percentage of the records that carry all the  items or contexts in the rule among those records that carry the items in the antecedent of the rule. A number of researchers \cite{mehrotra2016prefminer, srinivasan2014mobileminer, zhu2014mining, sarker2019recencyminer} have used association learning technique to mine rules capturing mobile phone users' behavior. However, it is well known that association learning technique produces a huge number of redundant rules \cite{fournier2012mining} that makes the rule-set unnecessarily large. In \cite{sarker2018mining}, Sarker et al. have shown that traditional association learning technique produces unnecessarily rules when applying on contextual smartphone data. Thus, it is very difficult for the decision making agents to determine the most interesting ones and consequently makes the decision making process ineffective and more complex \cite{bouker2012ranking}. 

Classification is another method that is frequently used in the area of machine learning and data science for solving the prediction problems. In general, classification is defined as a learning method that maps or classifies a data instance into the corresponding class labels that are predefined in the dataset. According to \cite{han2011data}, data classification is a two-step process; first one is the learning step where a classification model is constructed from a given dataset; the data from which a classification function or model is learned is known as the training set, and second one is a classification step where the model is used to test or predict the class labels for a separate unseen given data; the data set that is used to test the classifying ability of the learned model or function is known as the testing set. Several popular classification algorithms such as ZeroR, Naive Bayes, Support Vector Machines, K-Nearest Neighbors, Logistic Regression, Artificial Neural Network, Decision Tree, Random Forest have been proposed to build the prediction model \cite{han2011data}. 

Among these classification techniques, tree based context-aware model is more effective to predict mobile user activity in different contexts \cite{sarker2019classifications}. A very well-known and mostly discussed tree based technique for prediction is decision trees \cite{quinlan1986induction}. The core algorithm for building decision trees called ID3 proposed by J. R. Quinlan  \cite{quinlan1986induction}. ID3 algorithm constructs a decision tree by employing a top-down approach in which a greedy searching through the given training dataset is used to test each attribute or context at every node. It calculates the entropy and information gain which is a statistical property that is used to select which attribute to test at each node in the tree \cite{quinlan1986induction}. Based on the ID3 algorithm, a modified algorithm is proposed by Quinlan, namely C4.5 algorithm \cite{quinlan1993} builds decision trees from a training dataset in the similar procedure as ID3, using the concept of information gain. In particular, a number of researchers \cite{hong2009context} \cite{lee2007deploying} \cite{zulkernain2010mobile} \cite{sarker2019machine} \cite{sarker2017improved} have used decision tree classifier to model mobile phone users' behavior. Since we take into account our context-aware model as a multi-class problem that includes a variety of usage classes in a number of multi-dimensional contexts, a single decision tree based model may not be effective while predicting various categories of smartphone apps in different contexts. The reason is that a single decision tree may cause over-fitting problem, while selecting the root node based on contexts and consequently it may decrease the prediction accuracy of the resultant context-aware apps usage model \cite{sarker2019miim}.

Unlike the above approaches and context-aware models, in this work, we present a data-driven context-aware smartphone apps prediction model ``AppsPred'' using random forest machine learning technique that takes into account an optimal number of trees rather than a single decision tree based on relevant multi-dimensional contexts in order to make the model effective.

\section{Definitions and Problem Statement}
\label{Problem-Statement}
This section introduces main notions concerning individual's smartphone apps usage based on multi-dimensional contexts. In the following, the notion of smartphone apps usage dataset with relevant contexts, is formally stated.

\textbf{Definition 1.} \textit{Let $Con = \{con_1,con_2, ..., con_m\}$  be a set of contexts and $Q = \{q_1,q_2, ..., q_m\}$ the set of corresponding domains. A contextual smartphone apps usage dataset $DS$ is a collection of records, where -}
\textit{
	\begin{enumerate}[label=(\roman*)]
		\item each record $r$ is a set of pairs $(con_i,value_i)$, where $con_i \in Con$, and $value_i \in Q$. For example, if $con_i$ represents as the context `user mood', then an example of $value_i$ is `happy'. 
		\item each context $con_i \in Con$, also called attribute or contextual feature that may occur at most once in any record, and
		\item each record has a particular app usage of an individual user, e.g., using Microsoft Outlook.
	\end{enumerate}
}

\textbf{Definition 2.} \textit{Let, $Apps = \{App_1,App_2, ..., App_n\}$  be a set of smartphone applications that are used by an individual user $U$, each app $App_i$ represents a particular usage class for that user, which is taken into account in our multi-class context-aware problem.}

In this work, individuals' different categories of smartphones' apps such as social networking, communication, entertainment, or other daily life services related apps are taken into account in order to build an effective context-aware apps usage prediction model using machine learning techniques.

\textbf{Definition 3.}  \textit{Let $Con = \{con_1,con_2, ..., con_m\}$  be a set of contexts having influence on individuals to use different categories of smartphone apps mentioned above, according to their daily day-to-day situations, in the real world and $Q = \{q_1,q_2, ..., q_m\}$ the set of corresponding usage domains of an individual user $U$. Each context $Con_i$ and corresponding value can be used as a part of multi-dimensional contexts in our study, which can play a role to build a context-aware model according to the relevancy in apps usage.} 

Different contexts might have an influence on individuals in their daily life apps usages. For instance, an individual's apps usage behavior in her `happy' emotional state may be well different from her behavior when she is in a `sad' emotional state, which represents an example of user mood context. Similarly, other relevant contexts may also have the influence on their usages. As such, in this work, we take into account various types of relevant contexts, such as temporal context, spatial context, individual's mood and preference, work status, Internet connectivity, or device characteristics or status that might have influence on individuals' usage.

\textbf {Problem Statement.} With the above definitions, the main problem we are addressing in this paper is stated as follows:

\textit{Given, a smartphone dataset $DS$ containing different categories of apps usage history and corresponding contextual information of an individual mobile phone user. Our goal is to build an effective personalized context-aware apps prediction model based on the relevant multi-dimensional contexts related to individuals' day-to-day situations and preferences, and their devices' characteristics or status, in order to predict individuals' future usage for unseen test cases. In this paper, we present a data-driven context-aware apps prediction model ``AppsPred'' using random forest machine learning technique, for solving this problem.}

\section{Materials and Methods}
\label{Methodology}
In this section, we present our contextual smartphone apps usage datasets and the methodology for modeling personalized context-aware apps usage using machine learning techniques.

\subsection{Contextual Data Collection and Description}
In general, a context is defined as anything that can be used to characterize the situation of an entity \cite{dey2001understanding}. In this work, we take into account a number of contexts that have an influence on individuals to make a decision for using various categories of smartphone apps in their real-world life. These are: 

\textit{Temporal context:} It represents time related information. This is one of the primary context having influence on smartphone usage of an individual user \cite{sarker2017individualized}. For instance, smartphone apps usage of an individual in the morning might not be similar with her usage at night. Moreover, one's usage behavior may differ in different time periods or hours in the real world. Thus, we consider individuals' usage in each hour of a day while collecting the dataset.

\textit{Work status:} In general, individual's work status in the real world depends on day status like work day or holiday. Work status heavily impacts on apps usage for many individuals. For instance, one's apps usage behavior on Saturday, say a holiday, might not be similar with her usage on other work days.

\textit{Spatial context:} It represents users' spatial information like location, which can be treated as another significant context for modeling and predicting individual's smartphone apps. The reason is that phone usage of an individual can also be treated as a location based service \cite{sarker2017designing}. Thus, understanding user mobility and corresponding context-aware model is able to provide location based services for the benefit of individual users.

\textit{User mood:} Typically, user mood represents an emotional state of a human, which is mostly important related to sentiment and emotional analysis. As the emotional state of human being is not static in the real world, may change over time, it could be another significant context that impacts on individuals and to model personalized apps usage behavior. For instance, one individual typically likes to listen songs when she is in a happy mood, while likes to online messaging when she is in a sad mood.

\textit{Device status:} In addition to the above contexts related to users' day-to-day situations and preferences, individual's device related contextual information like phone profile, phone battery level or charging status might have an influence on individuals to use smartphone apps. For instance, if the phone battery of an individual's device gives low power signal, she typically is not interested to connect with the Internet for using an entertainment app.

\textit{Internet connectivity:} This also represents device related context that connects the device with the world. As such, Internet connectivity and speed might have an impact on individuals' smartphone usage. For instance, one individual likes to play video songs if Wifi (wireless fidelity that mainly refers to certain kinds of wireless local area networks) is available, otherwise not.

\begin{table*}[htbp!]
	\begin{center}
		\caption{An overview of contexts in our context-aware apps usage model}
		
		\label{tab:context-examples}
		\begin{tabular}{c|c|c} % <-- Alignments: 1st column left, 2nd middle and 3rd right, with vertical lines in between
			\textbf{Contexts} & \textbf{Type} & \textbf{Example values}\\
			\hline
			\makecell{Temporal} & Continuous & \makecell {Time [24-hours-a-day] \\ Day [7-days-a-week]}\\
			\hline
			
			\makecell{Work status} & \makecell{Categorical \\ (binary)} & \makecell {Holiday [yes, no]}\\
			\hline
			
			\makecell{Spatial} & Categorical & \makecell{User location [home, workplace, \\ canteen, playground, on the way, etc.]}\\
			\hline
			
			\makecell{User mood} & Categorical & \makecell{Emotional state [happy, sad, normal]}\\
			
			\hline
			
			\makecell{Device status} & Categorical & \makecell{Battery level[full, medium, low]}\\
			\hline
			
			\makecell{Phone profile} & Categorical & \makecell{Notification [general, silent, vibration]}\\
			\hline
			
			\makecell{Internet connectivity} & \makecell{Categorical \\ (binary)} & \makecell{WiFi status [on, off]}\\
		
			\hline
		\end{tabular}
	\end{center}
\end{table*}

Table \ref{tab:context-examples} gives a detailed picture of the contexts that are used in our context-aware model to predict personalized smartphone apps. We have collected smartphone apps usage datasets that include these contextual information from different individual users. All the participants in our study were university students and have their own smartphones. Data was collected from these participants from June 2018 to October 2018. The students have created an web interface for collecting these synthetic data from different users. The multi-dimensional contextual information discussed above and their interrelated patterns or relationships are of high interests to be discovered from the collected data, for the purpose of building a data-driven context-aware smartphone apps usage prediction model.

\subsection{Preprocessing of Contextual Data}
As we aim to build a machine learning based context-aware apps usage prediction model, we need exploratory data analysis to observe the characteristics of contexts. Thus, the first task for modeling is to make the apps usage dataset having multi-dimensional contexts able to feed our target machine learning classification technique. So for this reason, we first remove the missing data due to anomaly raised in contextual data collection. In order to build the context-aware model, we take into account all the relevant contexts that have an influence on individuals' usage as features, which is required in building a machine learning based model. Once the features have been identified, it is necessary to determine the features data type in the dataset. According to Table \ref{tab:context-examples}, all the contexts are categorical except temporal context. Thus, we take into account temporal segments with one hour interval in our analysis. In order to fit to the machine learning techniques, it is needed to convert all the categorical contextual features into vectors. The most popular approaches are ``Label Encoding'' and ``One Hot Encoding'' to do this task. In one hot encoding, the number of features increases with a significant number, and the resulting dataset will have lots of dimensions. On the other hand, in label encoding, the feature-values converted into a particular numeric number and the number of features remains the same. As we have taken into account multi-dimensional contexts, one hot encoded features might have sparse data which are difficult to fit in our target machine learning algorithm. Moreover, it takes a lot of processing time because of increased number of data dimensions. Thus, in this work, we use label encoding technique for converting the categorical contextual data into a feature vector to fit the contexts into the model. For instance, user mood can turn [happy, sad, happy, sad, normal] into [0, 1, 0, 1, 2] using label encoding.

\subsection{Contextual Random Forest Generation}
Once the preprocessing of contexts has been completed, we use machine learning techniques to build our context-aware apps usage prediction model. In order to build an effective context-aware model, we use random forest learning which is one of the most popular and powerful machine learning algorithms. A random forest learning \cite{breiman2001random} is an ensemble classifier that mainly combines randomly feature selection \cite{amit1997shape} and bootstrap aggregation \cite{breiman1996bagging}, in order to construct a collection of decision trees exhibiting controlled variation. 

To generate the random forest, we take into account all the relevant contextual features discussed above. The generated random forest consists a number of decision trees that can be used a separate classifier like a single decision tree based model. At each node in a tree, $d \ll D$ features are randomly selected from the D available features in the dataset, and the node is partitioned according to the Gini index \cite{breiman1984lclassification}. For a binary split, the Gini index of a node n can be expressed as - 

\begin{equation}
I_G(n) = 1 - \sum_{i=1}^{j}{p_i}^2
\end{equation}

where $p_c$ is the relative proportion of examples belonging to class $c$ present in node $n$. The best possible binary split is the one which maximizes the improvement in the Gini index.

\begin{equation}
\Delta I_G(n_p) = I_G(n_p) - p_1 I_G(n_l) - p_rI_G(n_r)
\end{equation}

where $p_l$ and $p_r$ are the proportions of examples in node np that are assigned to child nodes $n_l$ and $n_r$, respectively.

Finally, we combine the generated trees to form a single learner in order to produce final outcome. For a particular test case, it calculates the votes for each outcome predicted by each separate decision tree generated in the model and takes the highest voted predicted outcome, i.e., majority voting, as the final prediction result. For instance, we generate $N$ random decision trees utilizing the given training dataset. The terminal nodes of each decision tree represents apps usage behavior classes and the edges are associated with the corresponding contexts that are used for similarity matching in prediction. For a particular test case, each random decision tree generated in the model may predict different outcomes according to the contexts in the tree. In order to make the final prediction result for that contexts, we calculate and store all the predicted output for each tree and perform the majority voting among the $N$ trees. 

Since different values of the number of trees $N$ in random forest learning may give different prediction results, we determine an optimal value of $N$ based on low error rate or higher prediction results to build our target model. As more number of trees increases the overall computational cost, we take into account the lowest value of $N$ that gives higher prediction results in terms of $F_1$ score for a given training dataset, in order to establish the optimal value of $N$. Thus it needs to satisfy these two functions; $MIN(N)$ and $MAX(F_1 score)$, where $MIN$ and $MAX$ represents the minimum and the maximum respectively. Rather than arbitrarily assuming the number of trees $N$ in building the forest, we determine the optimal value of $N$ by iteratively varying the number. Thus, a random forest learning based context-aware model by generating optimal $N$ number of trees, is then considered as an effective apps usage model in terms of computational cost and prediction accuracy. 

\section{Evaluation and Experimental Results}
\label{Evaluation}	
To evaluate the effectiveness of our context-aware apps usage prediction model, we have conducted a range of experiments on the real mobile phone datasets collected by us. We have described about these datasets including the relevant contexts above. In the following, we report the experimental results utilizing these datasets and illustrate our AppsPred model with the detailed of experimental results of two users selected randomly.

\subsection{Evaluation Metric}
In order to measure the effectiveness of our AppsPred model in terms of prediction accuracy, we compare the predicted response with the actual response, i.e., the ground truth, and compute the accuracy in terms of:

\begin{itemize}
	\item Precision: It measures the ratio between the number of apps usage behaviors that are correctly predicted and the total number of apps that are predicted. If TP and FP denote true positives and false positives then the formal definition of precision is \cite{witten1999weka}:
	
	\begin{equation}
	Precision = \frac{TP}{TP + FP}
	\end{equation}
	
	\item Recall: It measures the ratio between the number of apps usage behaviors that are correctly predicted and the total number of apps that are relevant. If TP and FN denote true positives and false negatives then the formal definition of recall is \cite{witten1999weka}:
	
	\begin{equation}
	Recall = \frac{TP}{TP + FN}
	\end{equation}
	
	\item $F_1$ score: It is a measure that combines both the precision and recall defined above. It represents the harmonic mean of precision and recall. The formal definition of $F_1$ score is \cite{witten1999weka}:
	
	\begin{equation}
	F_1 \; score = 2 * \frac{Precision * Recall}{Precision + Recall}
	\end{equation}
	
	\item ROC value: It stands for Receiver Operating Characteristic (ROC). It can be another evaluation metric that also features on true positive rate, and false positive rate, to evaluate the machine learning classifier output quality. It summarizes the trade-off between true positive rate and false positive rate for a particular predictive model \cite{witten1999weka}.
\end{itemize}

\subsection{Experimental Results and Analysis}
To evaluate our AppsPred model, we employ the most popular K-fold cross validation technique in machine learning \cite{han2011data}, where we use $K = 10$, to measure the prediction accuracy. The 10-fold cross validation breaks the given dataset into 10 sets. It trains the prediction model on 9 sets and tests it using the remaining one set. This repeats 10 times and we take a mean accuracy rate. To show the effectiveness of each machine learning classification based model, we compare the accuracy, in terms of precision, recall, $F_1$ score and ROC value defined above. In the following, we report the overall experimental results in different dimensions.

\subsubsection{Personalized Prediction Results of our AppsPred Model}
In this experiment, we show the prediction results of our context-aware smartphone apps usage model for different individual users. For this, Table \ref{results-DS-01} and Table \ref{results-DS-02} show the prediction results in terms of Precision, Recall, $F_1$ score and ROC value, for each individual app as a usage class utilizing the dataset DS-01 and DS-02 respectively. As our AppsPred model is personalized, we show these results utilizing individual's datasets. If we observe Table \ref{results-DS-01} and Table \ref{results-DS-02}, we see that for each app usage class, the values of Precision, Recall, $F_1$ score, and ROC are significantly high, near to the maximum value 1, which ensures good prediction capability. This results also estimates that the FP rate representing instances falsely classified as a given class is ignorable. Thus, the overall experimental results in Table \ref{results-DS-01} and Table \ref{results-DS-02} show that our machine learning based context-aware model AppsPred is capable to effectively predict each app usage class of individual users according to their usage patterns in different contexts.

\begin{table}[htbp!]
	\centering
	\caption{The prediction results for various apps of a sample user using our context-aware model AppsPred utilizing dataset DS-01.}
	\label{results-DS-01}
	\begin{tabular}{|c|c|c|c|c|} 
		\hline
		Apps (Class) & Precision & Recall & $F_1$ Score & ROC value \\  
		\hline
		Gmail &  0.871   &   0.883  &   0.877   &   0.993 \\
		\hline
		Whatsapp  &  0.851  &   0.841   &   0.846    &  0.985 \\
		\hline
		Readnews  &  0.866  &   0.921   &   0.892   &  0.996 \\
		\hline
		LinkedIn    &  0.838  &   0.862   &  0.851   &    0.981 \\
		\hline
		Music &  0.852   &  0.848  &   0.851   &    0.991 \\
		\hline
		Youtube    &  0.878  &   0.878   &  0.878   &   0.992  \\
		\hline
		Facebook    &  0.832  &   0.839   &  0.835   &   0.987 \\
		\hline
		Skype       &  0.881   &  0.864   &  0.872   &   0.991 \\
		\hline
	\end{tabular}
\end{table}

\begin{table}[H]
	\centering
	\caption{The prediction results for various apps of a sample user using our context-aware model AppsPred utilizing dataset DS-02.}
	\label{results-DS-02}
	\begin{tabular}{|c|c|c|c|c|} 
		\hline
		Apps (Class) & Precision & Recall & $F_1$ Score & ROC value \\  
		\hline
		Facebook & 0.863   &  0.865  &    0.864   &   0.988 \\
		\hline
		Youtube  & 0.903  &   0.895   &  0.899   &   0.993 \\
		\hline
		Browser  & 0.855  &   0.875  &   0.865  &    0.991 \\
		\hline
		Gmail    & 0.887  &   0.913  &   0.902    &    0.994 \\
		\hline
		Whatsapp & 0.858  &   0.832   &   0.845  &    0.991 \\
		\hline
		Movie    & 0.901    &   0.857  &   0.878   &   0.987 \\
		\hline
		Games    & 0.862  &   0.881   &   0.871   &   0.991 \\
		\hline
		Live sports & 0.887  &    0.902  &   0.894  &    0.993 \\
		\hline
		Skype       & 0.911  &   0.883  &   0.897   &   0.991 \\
		\hline
		Instagram  &  0.902  &   0.909  &   0.905    &  0.994 \\
		\hline
		Read News  &  0.901    &   0.864   &  0.882   &   0.989 \\
		\hline
		LinkedIn   & 0.866  &   0.887   &  0.876   &   0.989 \\
		\hline
		Music      & 0.905  &   0.905  &   0.905   &   0.994 \\
		\hline
	\end{tabular}
\end{table}

\subsubsection{Effect on the Number of Trees}
\begin{figure*}[htbp!]
	\centering
	\begin{subfigure}[b]{.47\textwidth}
		\includegraphics[width=\textwidth]{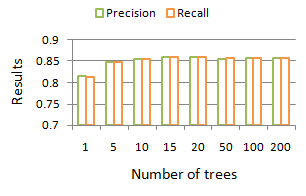}
		\caption{Prediction results (dataset DS-01).}
		\label{fig:comparison-D1}
	\end{subfigure}
	\begin{subfigure}[b]{.47\textwidth}
		\includegraphics[width=\textwidth]{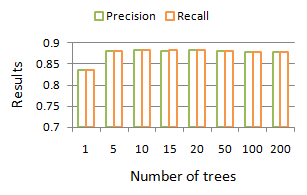}
		\caption{Prediction results (dataset DS-02).}
		\label{fig:comparison-D2}
	\end{subfigure}
	\caption{Prediction results in terms of precision and recall by varying the number of trees utilizing individual's datasets.}
	\label{fig:number-of-tree-effect}
\end{figure*}

In this experiment, we first show the effect on the number of trees on prediction accuracy utilizing individuals' apps usage datasets. To show the effect of the generated trees on prediction accuracy, we illustrate the detailed outcomes by varying the tree number for individual's dataset. For this, initially we consider one decision tree and the corresponding prediction results in terms of precision and recall defined above are measured. Figure \ref{fig:number-of-tree-effect} presents the impact of tree numbers on prediction accuracy (up to 200 decision trees) for different datasets DS-01 and DS-02 respectively. The x-axis of the figure represents the tree numbers and y-axis represents the corresponding prediction accuracy in terms of precision and recall, for the corresponding tree numbers for different datasets.

\begin{figure}[htbp!]
	\centering
	\includegraphics[width=.65\linewidth, keepaspectratio]{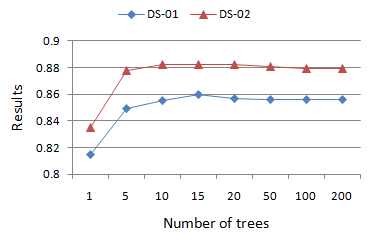}
	\caption{Effect on the number of generated trees on prediction results in terms of $F_1$ score for selecting the optimal number of trees.}
	\label{fig:optimal-selection}
\end{figure}

If we observe Figure \ref{fig:number-of-tree-effect}, we see that the prediction results are not static, it varies by varying the number of trees. As different number of trees give different prediction results, we determine the optimal number of trees based on $F_1$ score that combines the precision and recall. To do this, Figure \ref{fig:optimal-selection}, shows the effect on $F_1$ score by varying the number of trees for these datasets. The x-axis of the figure represents the tree numbers and y-axis represents the corresponding prediction accuracy in terms of $F_1$ score. 

If we observe Figure \ref{fig:optimal-selection}, we can see that initially the $F_1$ score is low, it increases up to a certain number of trees. The reason is that a single decision tree may cause over-fitting problem and gives lower $F_1$ score. On the other hand, multiple trees in random forest learning are generated from different subsets of data in our context-aware model, which control the problem of over-fitting and increases the $F_1$ score. As a result, it improves the prediction results that have been shown in Figure \ref{fig:optimal-selection}. If we observe more, we can see according to Figure \ref{fig:optimal-selection} that different number of trees give different $F_1$ scores. Thus, we select an optimal number of trees based on higher $F_1$ score with lower computational cost in terms of tree generation. As more number of trees increases the computational cost and make the model complex, we take into account that value as optimal for which it gives significant $F_1$ score with lowest computational cost. Thus, from Figure \ref{fig:optimal-selection}, we find that only 15 decision trees can produce significant result for dataset DS-01. Similarly, for dataset DS-02, an optimal number of decision trees is 10. These optimal number of trees make our AppsPred model simple and effective.

\subsubsection{Effect on the Execution Time}
In this experiment, we show the effect on the execution time of the number of trees utilizing individuals' apps usage datasets. To show the effect of generated trees on execution time, we illustrate the detailed outcomes by varying the tree number for individual's dataset. To do this, initially we consider one decision tree and the corresponding execution time is measured.

\begin{figure}[htbp!]
	\centering
	\includegraphics[width=.6\linewidth, keepaspectratio]{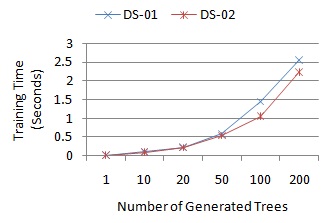}
	\caption{Effect of the number of generated trees on execution time.}
	\label{fig:time-effect}
\end{figure}

Figure \ref{fig:time-effect} shows the execution time taken by the context-aware model for different number of generated trees, starting from a single tree up to 200 trees, utilizing the dataset DS-01 and DS-02 respectively. The x-axis of the figure represents the tree numbers and y-axis represents the corresponding execution time, for the corresponding tree numbers for different datasets. From Figure \ref{fig:time-effect}, we see that if the tree size increases, it also increases the execution time that makes the model more complex. On the other hand, for small number of trees it performs efficiently.

\subsubsection{Effectiveness Comparison}
In this experiment, we show the effectiveness of our AppsPred model in terms of precision, recall, $F_1$ score, and ROC value, comparing it with some other popular classification techniques in machine learning. The comparing base methods are as follows:

\textit{ZeroR:} In the area of machine learning, this is the simplest approach for predictive analytics among the classification techniques \cite{witten2005data}. According to \cite{sarker2019classifications}, it can be used for deciding a standard execution as a benchmark for other classification techniques. For comparison purpose, we denote ZeroR leaning based model as BM1.

\textit{Naive Bayes (NB):} This in one of the most popular classification algorithms in the area of machine learning. A naive Bayes classifier \cite{john1995estimating} is a basic probabilistic based technique, which can foresee the class membership probabilities. For comparison purpose, we denote naive Bayes learning based model as BM2.

\textit{Support Vector Machines (SVM):} This is another popular classification technique used widely for various predictive analytics. In SVM \cite{keerthi2001improvements} a hyperplane is chosen in the vector machine, which is a line that can take part into the variable space. For comparison purpose, we denote support vector machines based model as BM3.

\textit{Logistic Regression (LR):} This is another popular probabilistic based statistical model used to solve the classification problems. Typically, logistic regression classifier \cite{le1992ridge} estimates the probabilities using a logistic function, which is also referred to as sigmoid function. For comparison purpose, we denote logistic regression learning based model as BM4.

\textit{Decision Tree (DT):} This is a very well-known and mostly discussed technique for classification and then used for predictive analytics. DT \cite{quinlan1986induction} constructs a decision tree by calculating the entropy and information gain which is a statistical property that is used to select which attribute to test at each node in the tree \cite{quinlan1986induction}. For comparison purpose, we denote decision tree learning based model as BM5.

\begin{figure*}[htbp!]
	\centering
	\includegraphics[width=.8\linewidth, keepaspectratio]{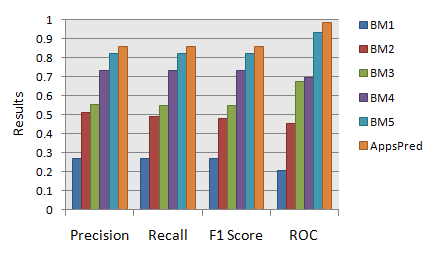}
	\caption{Effectiveness comparison with different classification based context-aware models utilizing a collection of datasets.}
	\label{fig:comparison-all}
\end{figure*}

To illustrate the effectiveness of our AppsPred model utilizing individuals' datasets, we show the relative comparison of precision, recall, $F_1$ score and ROC value. Figure \ref{fig:comparison-all} shows the comparison results in terms of average precision, recall, $F_1$ score and ROC value by considering a collection of datasets. For each classifier based approaches, we use the same training and testing set of data for the purpose of fair evaluation and comparison. If we observe Figure \ref{fig:comparison-all}, we find that our AppsPred model consistently outperforms other classification based methods for different datasets. In particular, the AppsPred model gives the highest prediction results in terms of precision, recall, $F_1$ score and ROC value. The reason for getting better result is that we take into account an optimal number of trees with subset of data and take the average result for the final outcome. Thus, it reduces the variance through averaging over learners, and the randomized stages decrease correlation between distinctive learners in the model. As a result, AppsPred is more effective than other classifier based approaches, when applying on mobile phone data consisting of variety of smartphone apps usage of individuals and corresponding multi-dimensional contexts.

\section{Discussion}
\label{Discussion}
The experimental results in Section \ref{Evaluation} have shown that our random forest learning based context-aware apps prediction model ``AppsPred'' is fully personalized and adaptive to individuals' usage behavior. Compared to the other popular classifier based approaches, the prediction accuracy in terms of precision, recall, $F_1$ score and ROC value, has been improved when this model is used, as shown in Figure \ref{fig:comparison-all}. Although it requires a number of iterations to determine the optimal number of decision trees to predict the future usage in a particular context, it is effective in terms of computational cost and prediction accuracy. The following are a few key discoveries from our study. 

\begin{itemize}
	\item To predict individuals' apps usage behavior in a particular context, random forest learning based apps prediction model having an optimal number of decision trees is more effective than a single decision tree based model. In our experiments, we have shown the corresponding results in terms of precision, recall, $F_1$ score and ROC value utilizing individuals' datasets.
	
	\item Another important finding of our study is that a large number of trees in random forest learning based context-aware model is not always effective in terms of prediction results and computational cost. According to Figure \ref{fig:number-of-tree-effect} and Figure \ref{fig:optimal-selection}, for a single tree the prediction accuracy in terms of precision, recall and $F_1$ score is low, it increases up to a certain number of trees. After that, although it increases the computational cost in generating more trees but no significant prediction results are found.
	
	\item We have observed a significantly lower prediction accuracy when using other classification based approaches compared to our AppsPred context-aware model. The reason is that other models cannot capture the different categories of apps usage patterns properly in multi-dimensional contexts. Consequently, these approaches have low prediction accuracy while comparing with AppsPred model that uses random forest learning with optimal number of trees to capture the usage patters more properly.
	
	Overall, our context-aware model AppsPred is more effective according to its prediction results with less computational cost. Although it takes higher training time than a single decision tree based model, it shows the effectiveness in terms of prediction accuracy better than a decision tree based model. Typically, a machine learning algorithm searches a space of hypotheses to find the best hypothesis for a particular problem. For a small amount of training data, a learning algorithm could identify various hypotheses providing the similar accuracy on the testing data. An ensemble of a number of single classifiers can average their prediction results and thus avoid selecting the inaccurate classifier. Thus, this random forest learning based context-aware apps usage model AppsPred is very helpful for predicting individuals' future usage in a particular context-aware test case.
\end{itemize}

\section{Conclusion}
\label{Conclusion}
In this paper, we have presented a data-driven context-aware smartphone apps usage model AppsPred that utilizes random forest machine learning technique by taking into account an optimal number of decision trees. In order to build this personalized model, we have collected apps usage datasets from individual users and takes into account the relevant contextual features that have an influence on individuals for using various categories of apps. No assumption or prior  knowledge is needed in employing our model as we select the optimal number of trees dynamically according to the data patterns, which may vary from user to user. Experimental results on the collected contextual smartphone datasets indicate that our model outperforms popular classifier based models for predicting individuals' smartphone apps. In the paper, we have demonstrated that our AppsPred model is in general highly effective in terms of accuracy and computational cost in predicting future usage with multi-dimensional contexts. We believe that this model will be helpful to application developers to build corresponding real-life applications, in order to provide context-aware personalized services for the end users according to their needs. To assess the effectiveness of AppsPred model in application level by conducting a user survey, could be a future work.

\section*{Acknowledgment}
\label{Acknowledgment}
The authors would like to thank all the participants, who are involved in this study for collecting their smartphone apps usage datasets consisting of various categories of apps and corresponding contextual information.

\section*{References}
\bibliographystyle{plain}
\bibliography{AppsPredbib}

\end{document}